# Intellectual Stewardship:

## Re-adapting Human Minds for Creative Knowledge Work in the Age of AI

**Note: This is a preprint.*

Jianwei Zhang
Department of Educational Theory and Practices,
University at Albany, State University of New York, Albany, NY
jzhang1@albany.edu

### Abstract

**Background:** Amid the opportunities and risks introduced by generative AI, learning research needs to envision how human minds and responsibilities should re-adapt as AI augments or automates various tasks and enters daily learning, knowledge work, and social life.
**Approach:** Drawing on theories of learning, intelligence, and knowledge creation, this conceptual paper proposes intellectual stewardship as a human-centered framework that foregrounds human wisdom and judgment for advancing creative learning practices with AI.
**Key points:** Students and teachers work as responsible governors of intellectual processes distributed across human and artificial systems, guided by five core principles. Being *knowledge-wise* involves understanding the evolving state of knowledge and taking purposeful actions to advance it. Being *intelligence-wise* emphasizes making informed choices about how to orchestrate distributed cognitive processes and resources. Being *context-wise* requires sensitivity to opportunities and risks in changing environments. Being *ethics-wise* foregrounds ethical judgment, responsibility, and care in the use of knowledge and intellectual power. Finally, *self- and community-growing* defines the overarching purpose, aligning intellectual work with personal development and the advancement of collective well-being.
**Contribution:** The principles provide a lens for viewing the adaptation of human minds in AI-infused learning environments, calling for the development of meta-level dispositions and capabilities that characterize wisdom-oriented, socially responsible knowledge builders in the AI age.

## Introduction

As generative AI (GenAI) transforms how people process information, solve problems, communicate, and get work done across social sectors and industries, education faces an urgent demand to develop AI-ready learners, citizens, and future workers (McKinsey & Company, 2025; Miao & Holmes, 2023; U.S. Department of Education, 2023; World Economic Forum, 2025). So far, dominant uses of AI in education have centered on optimizing student mastery of predefined knowledge and skills within existing instructional frameworks. This conceptual paper seeks to advance a different vision: harnessing AI to enable more creative and impactful learning for all learners, who work as collaborative knowledge builders to investigate authentic challenges and advance knowledge of value to their communities and the broader world (Bereiter & Scardamalia, 2025; Linn et al., 2025; Paavola & Hakkarainen, 2005; Penuel et al., 2025; Resnick, 2024).

While AI technologies have shown a compelling potential to support creative knowledge work, they also intensify concerns about student agency, responsibility, integrity, and safety (Burns et al., 2026; Kasneci et al., 2023; Wu et al., 2025). Much of the educational response has focused on improving AI *use*, framing AI as tools to be used or risks to be mitigated. What is missing is a broader, coherent account of *what* and *how* people should learn in the age of AI. Specifically, what mindsets are essential, and how should human minds and responsibilities re-adapt as AI augments or automates cognitive tasks and enters daily learning, knowledge work, and social life? This paper proposes *intellectual stewardship* as such a framework, positioning learners and knowledge workers as governors of intellectual processes performed by human, AI, or hybrid systems. Specific choices about AI use are embedded in the broader context of creative knowledge work, integral to meta-level decisions about what knowledge matters, how to advance it, for what personal, social, and ethical purposes. I first review the related literature, then introduce the concept of intellectual stewardship and propose five core principles that lay out epistemic stances and decision points for directing distributed intellectual processes in AI-infused environments.

## What New Opportunities and Challenges Are Emerging?

Scholars recognize the potential use of GenAI to augment learners' creative knowledge work: to explore complex problems, test ideas, and engage in sustained inquiry that would otherwise be difficult to implement (Markauskaite et al., 2022; Siemens et al., 2022). GenAI may act as a conversational partner, critique generator, or resource explorer in collaborative learning environments. A line of research builds on the Knowledge Building theory and pedagogy, which conceptualizes learning as integral to personal and collective efforts to advance the community's knowledge (Chan & van Aalst, 2018; Scardamalia & Bereiter, 2014). The findings show that GenAI may help students formulate productive questions, generate candidate explanations, explore alternative framings, compare and synthesize sources, enrich collaborative discourse moves, and support progressive questioning that sustains ever-deeper inquiry (Ba et al., 2025; Bereiter & Scardamalia, 2016, 2025; Feng, 2025; Lee et al., 2023; Yi, Yin, & Zhang, 2025). For example, Chen, Zhu, and Díaz del Castillo (2023) examined how students used ChatGPT across different stages of a knowledge-building inquiry. The analyses identified productive patterns of student–ChatGPT interaction: to support problem definition, idea generation and evaluation, critical discourse, and conceptual rise-above. The teacher and students worked together to manage AI's limitations through fact-checking outputs and refining prompts.

Alongside these opportunities, the literature documents significant risks. First, cognitive offloading has emerged as a serious concern. Unreflective AI use may shortcut personal thinking and reduce the mental effort necessary for deep learning (Shen & Tamkin, 2026). Kosmyna and colleagues (2025) report neurocognitive evidence that heavy LLM reliance is associated with reduced deep semantic encoding and weakened ownership over written work — what they call "cognitive debt". This finding converges with a meta-analysis showing that AI-driven performance gains may not translate into lasting learning (Deng et al., 2025).

Second, overreliance on AI may erode student epistemic agency. GenAI introduces a symbiotic learning partnership that can either strengthen or erode human epistemic agency, depending on learners' epistemic stances (Wu et al., 2025). Many young learners are prone to accepting AI-generated contributions without sufficient evaluation or refinement. Research shows that while GenAI use can enhance student idea generation and task completion, overreliance on AI systems may reduce student metacognitive effort, weaken their critical thinking and ownership over ideas, and foster dominant thinking patterns at the expense of idea diversity (Choi et al., 2025; Fan et al., 2025; Gerlich, 2025; Habib et al., 2024).

Third, GenAI raises ethical and civic risks. By enabling rapid production of persuasive content at scale, AI amplifies the spread of misinformation, bias, and misuse of AI-generated material, threatening academic integrity, social trust, and student well-being (Burns et al., 2026). Ethics guidelines improve governance and accountability (European Commission, 2024; UNESCO, 2021), but in education, ethical judgment cannot be an external add-on; it must be embedded in the social practices and tools through which learning takes place.

## Searching for a Conceptual Framing

Educators face a critical crossroads as they determine how to move forward amid the potential benefits, risks, and uncertainties associated with AI. Responses of educators vary widely, from enthusiastic hype to AI fear (Zhai & Krajcik, 2024). So far, education research has focused on developing practical strategies to guide AI use, including explicit teaching of prompt design and critical evaluation of AI outputs. Increasing attention has been devoted to AI literacy: "the knowledge and skills that enable people to critically understand, evaluate, and use AI systems and tools to safely and effectively participate in an increasingly digital world" (Mills et al., 2024, p. 4). What is missing is a coherent educational framing that situates specific tool uses within a broader theory of learning in the age of AI (cf. Markauskaite et al., 2022). What mindsets must students develop to engage with AI productively as creative learners and knowledge builders, avoiding the risk of "smart people stumbling at their own smartness"?

This paper intends to address this need, building on the literature on human learning, intelligence, and AI research that highlights human–AI partnerships and human-centered AI. The concept of cognitive partnerships between humans and technologies is foundational to computing research (Engelbart, 1963). With the rise of GenAI, scholars have extended this line of thinking to examine how human–AI partnerships may reshape learning and intellectual activity (Siemens et al., 2022; Markauskaite et al., 2022). Human–AI partnerships refer to the systemic arrangements through which intellectual work is distributed and coordinated across humans and artificial agents in an authentic context of knowledge practices. Intelligence emerges from dynamic interactions between the human minds (i.e., internal cognitive structures and processes) and the people, knowledge artefacts, technologies, and cultural practices in the larger (wild) world (Hutchins, 1995; Pea, 1993; Perkins et al., 2000). This concept is valuable for analyzing how humans and AI collaborate to generate synergistic outcomes. For example, in knowledge co-creation, AI may be directed to play various roles: a subcontractor of specific tasks, critic of human-generated works, or teammate for co-thinking. Different roles imply different allocations of cognitive initiative, control, and responsibility (Lin & Riedl, 2023).

Complementing this perspective, human-centered AI (HCAI) emphasizes designing and using AI systems in ways that augment human capacities while maintaining human control, reflecting human values and ethics, and improving human well-being (Schmager, Pappas, & Vassilakopoulou, 2025; Stanford HAI, 2022). This principle is particularly important for education. In the age of AI, "it is more important than ever for children from diverse backgrounds to…develop the most human of their abilities—the abilities to think creatively, engage empathetically, and work collaboratively—so that they can deal creatively, thoughtfully, and collectively with the challenges of a complex, fast-changing world" (Resnick, 2024, p.2).

Accordingly, the design of human–AI partnerships for learning must foster student epistemic agency: their personal and collective capacity to actively shape the knowledge-building work for consequential outcomes, including knowledge goals, processes, tools, and collaborative roles (Damşa et al., 2010; Miller et al., 2018; Scardamalia, 2002; Wu et al., 2005; Zhang et al., 2022). While GenAI excels at surfacing patterns across large bodies of information, humans remain responsible for the reflective functions that define epistemic agency (Woodruff & Hewitt, 2026). This means AI-supported environments must encourage learners to regulate their learning, generate questions, pursue deep understanding, and refine AI contributions rather than passively accept them (Chen, 2025; Järvelä et al., 2023).

Building on these concepts, this paper aims to develop a more nuanced, human-centered framework to explain how learners enact epistemic agency and judgment to configure dynamic knowledge-building practices in an AI-infused environment (Pendleton-Jullian & Brown, 2018; Tao & Zhang, 2021; Zhang et al., 2022). The term human-AI partnership captures the structural arrangement well, yet it says little about the essential reflective functions of the human mind in governing that interaction with agency. The metaphor of stewardship is proposed to capture such high-level governing functions. In what follows, I flesh out the framework of intellectual stewardship that foregrounds human epistemic responsibility and ethical judgment in harnessing intellectual power for advancing knowledge and growing personal and social good.

## Intellectual Stewardship

The motif of the wise steward recurs across cultures, religions, and historical periods. In ancient China, Guan Zhong (725–645 BC), appointed Prime Minister of the state of Qi after imprisonment for backing the wrong prince, reorganized markets and armies, turned liabilities into revenue, and counseled alliance over conquest, never mistaking the state's resources for his own. His story echoes that of Joseph in the Bible, who rose from slavery and imprisonment to become Prime Minister of Egypt, storing a fifth of every harvest during seven years of abundance so that when famine struck the whole region, the storehouses held. The wise stewardship they represent offers a productive metaphor for understanding human wisdom and responsibility in AI-infused knowledge work. A steward, whether of a small entity or a larger enterprise, is entrusted with resources that do not ultimately belong to them and is therefore responsible for managing those resources wisely and fruitfully. Unlike a passive caretaker or an unchecked owner, a steward exercises judgment over how resources are allocated, coordinated, and developed so that the whole enterprise flourishes. This involves trustworthiness and accountability, discernment of present needs and future challenges, and a commitment to growth — not mere preservation — through thoughtful planning, sustained effort, and responsible risk-taking. Therefore, wise and faithful stewardship demands high-level, thinking-intensive efforts, no less than ownership. At the core of such stewardship is a mindset of reflective judgment, thoughtfulness, and diligence that goes *above and beyond* task completion: staying metacognitively above one's work with agency and intentionality, making sense of dynamic changes in context, and approaching every task as an opportunity for meaningful growth and consequential change in the near and longer terms. As Guan Zhong himself said: "A year grows grain; a decade grows timber; a century grows a people."

Recent research shows that productive knowledge workers tend to practice a form of AI stewardship: turning intentions into queries, shaping AI responses, and taking responsibility for the outcomes (Lee et al., 2025). Good AI stewardship requires a broader view of the knowledge practices that AI serves. I propose the concept of *intellectual stewardship* to capture this broader view. Intellectual stewardship refers to the human practice of wisely governing intellectual activities--whether human, artificial, or hybrid--by judging what knowledge work should be undertaken, how it should be pursued, with what tools and resources, and toward what personal and social purposes. The stewardship metaphor provides a lens for analyzing human agency and responsibilities: What assets and practices fall under one's stewardship? What critical judgments need to be made, with whom, in what contexts, and for what purposes? What principles guide reflective judgment toward fruitful outcomes? *Intellectual* stewardship further centers epistemic agency and responsibility for orchestrating creative knowledge work using various intellectual tools within a collaborative community. With a stewardship mindset, knowledge workers approach knowledge as a personal and social asset to continually grow, not merely information to be stored and consumed. Specific decisions about AI use — which tools to employ, when, in what roles, and with what risks — become integral to their broader commitment to continually adapt and advance their knowledge-building practices.

Applying this metaphor to education, students are intellectual stewards in apprenticeship. They are entrusted with opportunities and spaces to learn, teachers and peers to work with, rich knowledge resources to build upon, and AI and other technologies to use. As they develop their own knowledge and capabilities, students also assume shared responsibility for advancing the knowledge and well-being of their learning community and for contributing to broader communities where possible. They enact epistemic agency by participating in high-level decisions about what to inquire into, why particular problems matter, and how collaborative inquiry should be organized over time (Zhang et al., 2018, 2022). As AI catalyzes deeper changes in curriculum, assessment, and school management, future education should give students greater flexibility to orchestrate learning resources and activities in pursuit of personal interests and creative goals (Penuel et al., 2025).

## How Do Students Enact Intellectual Stewardship?

This section further unpacks how students enact intellectual stewardship, examining what kinds of intellectual assets and practices fall under their stewardship, what critical judgments are to be made with their teacher and peers, in what contexts, for what purposes. Students steward two core forms of intellectual assets: (a) *content knowledge* across disciplinary domains as related to personal experiences, and (b) *intelligent capacity and tools* for processing and advancing knowledge. Both are substantially augmented by AI technologies. Related to these assets, time is an essential and finite resource under student stewardship. AI stands to reshape how time is *invested*, not merely *spent*. As routine cognitive work is offloaded to AI, the risk is passivity; the opportunity is reinvestment. Gains in efficiency should free time and attention for higher-value activities: tackling more complex problems, pursuing new lines of inquiry, building character and passion, and assuming greater responsibility for community knowledge and well-being. The central challenge is learning to make this reinvestment intentionally, so that students--and teachers--continually adapt and surpass themselves in a changing world (Bereiter & Scardamalia, 1993; Schwartz, 2025).

To guide wise and fruitful intellectual stewardship, I propose five core principles that highlight the reflective judgments human minds need to prioritize in AI-infused learning environments (see Figure 1). Students, as intellectual stewards, must learn to be:

1. **Knowledge-wise**: understanding the evolving state of knowledge and taking purposeful actions to continually advance it with AI support;
2. **Intelligence-wise**: making wise choices to orchestrate distributed intellectual processes and resources;
3. **Context-wise**: recognizing opportunities and risks in a changing environment to develop creative contributions and responsive actions;
4. **Ethics-wise**: exercising ethical discernment, responsibility, and care in the use of knowledge and intellectual power; and
5. **Devoted to self- and community-growing**: continually growing and surpassing themselves while advancing the collective good of their communities and the broader society.

The first four principles define distinct facets of choice-making; the fifth highlights the human-centered purpose toward which the choices are directed. Together, they provide a compass for navigating complex learning practices in AI-infused environments. I unpack each principle in turn, addressing what it means, why it matters, the meta-level mental functions involved, and the specific stances and judgment points it calls for.

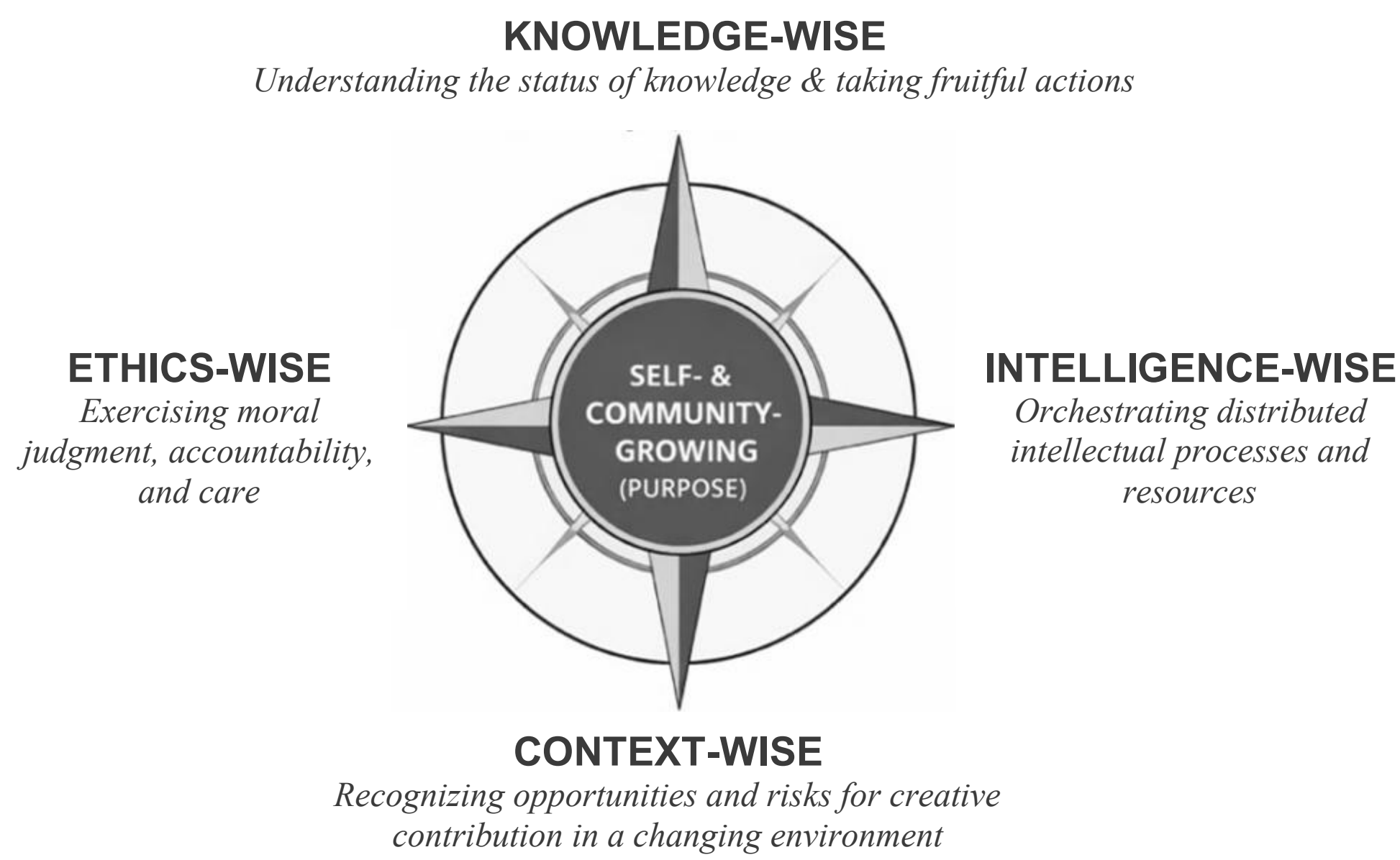

Figure 1. Five principles of intellectual stewardship for governing AI-mediated knowledge work.

### *Knowledge-Wise*

Beyond learning to be more *knowledgeable,* students in AI-infused environments need to develop a *knowledge-wise* mind: a mindset that can exercise reflective judgment about the knowledge they engage with—its epistemic status, explanatory power, limitations, gaps, and future potentials—in order to act strategically to advance knowledge and put it to productive use.

Under their stewardship, students work with both personal knowledge and collective knowledge. Personal knowledge refers to their formal knowledge gained through education and informal experiences accumulated in life. Collective knowledge is what Popper (1972) termed *World 3*, the realm of objective knowledge artifacts such as theories, explanations, arguments, designs, and models that exist beyond individual minds and are open to public critique and improvement. Frontline research reframes students' relationship to collective knowledge: they are not only knowledge users or recipients, but also knowledge builders who participate in the advancement of collective knowledge (Paavola & Hakkarainen, 2005; Scardamalia & Bereiter, 2014). As knowledge-wise stewards, they need to develop a reflective sense of the evolving ideas in their own learning community, tap in the knowledge of the larger fields, and seize on opportunities to make creative contributions.

Such knowledge-wise stewardship becomes essential for students to use AI to advance knowledge while mitigating its epistemic risks. Trained on vast corpora of collective knowledge, GenAI systems are reshaping the ecology of collective knowledge itself, changing how people encounter, generate, synthesize, and circulate knowledge. This has both positive and negative implications. On one hand, GenAI may widen access to World 3 by helping students interpret complex texts, compare different perspectives, work across different modalities and media, and bridge linguistic barriers, opening opportunities for students to participate in knowledge-creating discourse (Bereiter & Scardamalia, 2025; Feng, 2025; Vokatis et al., 2026). At the same time, AI amplifies epistemic risks (Hou et al., 2026; Kosmyna et al., 2025). Easy access to AI outputs may encourage surface-level engagement and premature conclusion, weaken learners' responsibility for deep thinking and judgment, and accelerate the spread of underdeveloped, biased, or misleading knowledge artifacts in the knowledge ecology. The knowledge-wise orientation is essential to guiding students' reflective judgment and action in real contexts, so they can capitalize on the affordances of AI while addressing the potential risks in a timely and responsible manner.

Central to knowledge-wise stewardship is students' epistemic cognition and meta-knowledge. Epistemic cognition refers to people's beliefs, theories (mental models), and dispositions related to knowledge and knowing, such as the aim and process of knowledge work and criteria of success (Barzilai & Chinn, 2018; Chinn & Sandoval, 2018). As a key part of epistemic cognition, meta-knowledge refers to knowledge about the state of knowledge. It helps learners to "see the forest through the trees" as they navigate their unfolding paths of knowledge building, situating specific works and ideas in a larger problem space that matters to them (Bereiter & Scardamalia, 2025). Epistemic cognition and meta-knowledge function as a form of intellectual governance in AI-infused environments, guiding student reflection on key issues about their knowledge work. Example decision points include: which problems are worth pursuing, what information is relevant, which ideas are promising to develop (or dead-ends to avoid), what/how AI systems should be used to advance the ideas and bring in new perspectives, whether AI outputs advance the understanding of the central problems or merely add rhetorical refinement, what problems have been addressed, what remains unclear, and how to move forward. Engaging in such decision-making, knowledge builders enact high-level responsibility to make sense of the landscape of the knowledge in their community and position AI's role in this context. As they reflect on their knowledge-building process with AI, learners may frame/reframe problems, prioritize inquiry directions, coordinate personal and collaborative efforts, and take responsibility for the trajectory and integrity of knowledge advancement. AI outputs themselves become objects of inquiry subject to evaluation, revision, and continual improvement.

As a key epistemic marker of knowledge-wise stewardship, students need to adopt an expansive, agentic stance to continually build on what they know and push beyond: asking deeper questions, identifying gaps, connecting ideas and inquiries across contexts, and launching new lines of inquiry driven by a sense of wonderment and love of knowing (Engle et al., 2012; Vokatis et al., 2026; Wu et al., 2025; Zhang et al., 2022). Such intentionality may direct student-AI interaction to pursue ever-deeper inquiry, leading to increasingly deeper understandings, better solutions, and higher-level creative expertise over time (Bereiter & Scardamalia, 1993; Sawyer, 2015; Schwartz, 2025).

### ***Intelligence-Wise***

To work with AI productively, students not only need solid intelligent skills but also learn to be *intelligence-wise*, that is, knowing how to use their capabilities and skills wisely in collaboration with human and AI partners for productive outcomes. This orientation guides how learners work with their personal and distributed intelligent capabilities under their stewardship. Personal abilities include diverse ways of thinking—analytical, creative, and practical (Sternberg et al., 2023)—that help learners to process knowledge, solve problems, and make sound decisions. Their personal abilities are further extended by distributed intelligence systems, which involve interacting with other people, tools, representations, and social practices in authentic contexts (Hutchins, 1995; Pea, 1993). As AI reshapes the distributed cognitive systems, education must cultivate human minds that know how to govern different intelligent capacities and processes, including how different human thinking approaches and AI-based operations are selectively used and synergized to achieve productive and ethical purposes.

Intelligence-wise stewardship is critical to AI-supported knowledge work. Rather than merely assisting human cognition, GenAI increasingly reorganizes how cognitive work is allocated and coordinated across human and machine systems. Serious risks come with this shift: cognitive offloading without understanding, weakened epistemic agency, and misalignment between task-level performance and deep learning (Deng et al., 2025; Grinschgl & Neubauer, 2022; Kosmyna et al., 2025; Riedl et al., 2024; Wu et al., 2025). To mitigate such risks, it is essential for learners to exercise reflective judgment over when AI is used and how the distributed cognitive processes are coordinated and constrained.

Intelligence-wise learners engage in meta-level thinking about how human and AI contributions should be leveraged and orchestrated to achieve meaningful goals. Key decision points address issues such as: what kind of intellectual work is needed in a context, for what purposes, how should it be carried out, by/with whom, using what intellectual resources and tools, and what counts as good work or success? By engaging in reflective judgment and regulation of these issues, students build reflective framing of their inquiry-based work, which guides personal and collective efforts to advance their knowledge (Zhang et al., 2018, 2022). Guided by their framing and needs, they may guide AI agents to play specific roles as thinking partners, such as a divergent thinker, questioner, or synthesizer.

A core mental function enabling intelligence-wise stewardship is *meta-intelligence*: a distinctly human capacity for governing how different intellectual processes and resources are mobilized for specific purposes and contexts. The concept of meta-intelligence was originally proposed in relation to individual problem solving (Sternberg et al., 2023), defined as higher-order processes that select, control, and coordinate analytical, creative, practical, and wisdom-based approaches in a problem situation. These processes include recognizing and framing problems, allocating mental resources, monitoring problem-solving strategies, and evaluating outcomes. In AI-infused environments, the concept of meta-intelligence needs to be expanded beyond personal thinking. Meta-intelligence serves as a high-order governing ability over distributed human–AI systems, enabling learners to frame intellectual work in context, orchestrate cognitive processes across people and machines, align intellectual efforts with values and purposes, and assume responsibility for the consequences of their intellectual work.

As key markers of meta-intelligence, education needs to nurture productive dispositions toward thinking with AI. Research on the dispositional view of intelligence highlights a constellation of stances (habits of mind) that shape how learners engage with intellectual work, potentially applicable to working with AI (Lucas, 2016; Perkins et al., 2000; Woodruff & Hewitt, 2026; Wu et al., 2025). Inquisitive and persistent dispositions may orient students toward identifying valuable problems and sustaining effort to achieve deep understanding that goes beyond AI-generated first responses. Open-minded yet disciplined dispositions support the imaginative exploration of possibilities surfaced through AI interaction while maintaining standards of evidence, coherence, and logical reasoning. Collaborative dispositions enable learners to engage in productive dialogue with both human peers and AI partners—generating diverse ideas for creative exploration, examining multiple perspectives on shared problems, questioning underlying assumptions, connecting and synthesizing ideas across sources, and monitoring the progress of inquiry while identifying new problems to pursue (Bereiter & Scardamalia, 2025; Chen et al., 2023; Lee et al., 2023). Students and teachers need to be mindful about the limitations and fallibility of both human and artificial intelligence, so results and ideas are critiqued, risks are mitigated, and problems are addressed.

Intelligence-wise and knowledge-wise orientations address complementary dimensions of human responsibility: the former governs how thinking processes are organized, resourced, and adapted, while the latter governs how conceptual ideas are developed, evaluated, and advanced. The integration of both guides the fruitful use of intellectual power for advancing deep understanding and inquiry, focusing on issues that are meaningful to students in an authentic context.

### ***Context-Wise***

Context matters when it comes to how knowledge and intelligence are used and advanced. In AI-infused environments, learners need to be *context-wise* — capable of positioning their intellectual activities in relation to an evolving situation. They need to immerse themselves in authentic settings of creative work and human experience, recognizing emerging needs and opportunities to take intellectual initiatives in connection with the ideas of others and information generated by AI systems.

Context-wise stewardship is critical in the age of AI. A core limitation of AI systems is their lack of lived, embodied, and culturally situated experience; they cannot fully grasp the significance of unfolding events, social meanings, or moral stakes embedded in real-world contexts (Bereiter & Scardamalia, 2025). Context-wise stewardship foregrounds human responsibility for interpreting situations and governing when, how, and why AI-mediated intellectual work should proceed, shifting attention from optimizing responses within a predefined task frame to judging which problems and directions are worth pursuing in the first place. This extends beyond context-aware AI design, which focuses on embedding learner characteristics into systems to improve AI responsiveness (Liu et al., 2025), toward something more fundamental: human judgment over the meaning and significance of context itself.

This orientation builds on dispositional and situated views of intelligence and creativity (Pendleton-Jullian & Brown, 2018; Perkins et al., 2000; Sawyer, 2015), which emphasize sensitivity to opportunity as a core component of dynamic intelligent action. In AI-infused knowledge environments characterized by rapid change, ambiguity, and interdependence, students must learn to attune themselves to emerging problems and opportunities, connect with people holding diverse interests and expertise, and take responsive action to develop productive paths of inquiry. Our studies of opportunistic collaboration in Knowledge Building classrooms illustrate such dynamic learning (Zhang et al., 2009, 2022): rather than working in fixed groups with prescribed roles, students formed and reformed inquiry groups in response to evolving questions and ideas, continually building on one another's observations, incorporating new resources, and generating spontaneous inquiry activities to advance collective understanding. On an ongoing basis, they read emerging currents of ideas and improvised collaborative work as conditions shifted.

To cultivate context-wise stewardship, schools need to create authentic learning environments that support spontaneous participation and interaction. Learners attend to tensions and advances in collaborative dialogue, identify emerging knowledge connections and gaps, and connect their inquiry to broader disciplinary and social developments. They recognize unexpected opportunities — new questions sparked by a peer's idea, unforeseen connections surfaced through human or AI interaction, or deeper directions emerging from exploratory work — and intentionally seek new challenges as progress is made. Teachers, similarly, need to increase flexibility and responsiveness, engaging in idea-centered noticing as student inquiry unfolds: seeing ideas under construction, envisioning opportunities for deeper thinking, and responding in ways that help students grow their ideas together (Park & Zhang, 2025). AI-powered analytics may help students and teachers navigate dynamic flows of ideas, pointing toward emerging directions for deeper inquiry (Chen & Zhang, 2016). Students bear ultimate responsibility for deciding where to direct attention, when to reframe or expand inquiry, and how to reorganize their personal and collaborative work over time.

### *Ethics-Wise*

Such stewardship of AI-infused environments calls for an *ethics-wise* mind, which reflects on ethical and moral responsibilities to others, to the knowledge and well-being of their learning community, and to the broader communities and society in which their intellectual work is embedded. Ethical issues become especially salient when working with AI, which scales intellectual power while weakening the processes that ordinarily slow down our thinking, invite verification, and reinforce personal responsibility. This combination increases the likelihood that errors, bias, misrepresentation, and even harmful manipulations (e.g., of personal images) circulate through classrooms and public spaces. Intellectual stewards must exercise ethical judgment, care, and accountability.

Research on AI literacy frameworks and policies provides guidelines for users to make responsible use and critical evaluation of AI, attending to risks such as misrepresentation, bias, erosion of responsibility, and harm to individuals, communities, and knowledge systems (Mills et al., 2024; UNESCO, 2021). In practice, educational institutions are also creating their own AI policies and rules. Yet ethical stewardship extends beyond rule-compliant user behaviors to an *ethical mind,* which is intrinsic to creative knowledge work itself, with or without AI. AI technologies are rapidly evolving, so too are the ethical landscapes around AI use. This makes it almost impossible to develop comprehensive rules and guidelines. For this reason, schools need to shift their focus to cultivating a principle-based, ethical mindset among students and educators to guide their discernment in specific contexts. Drawing on Gardner's (2006) framing, the ethical mind operates at a reflective level, guiding how individuals understand and enact their roles as learners, workers, professionals, and citizens. When working with AI, ethics-wise minds do not ask only whether something *can* be done efficiently with AI, but whether it *should* be done, with what potential benefits and risks to themselves and to others, and how to safeguard with care.

At the core of the ethical mind is a commitment to "good work," which integrates intellectual/technical excellence, personal meaning, and ethical integrity and responsibility (Gardner, 2006). Good work for learning and knowledge building is not only about productivity; it needs to be ethical in the first place. Therefore, a steward with an ethical mind would strive to advance good work that benefits personal and collective good while addressing moral failures with responsibility and integrity. GenAI can rapidly produce authoritative-sounding explanations and claims, increasing the risk that underdeveloped, biased, or misleading ideas enter shared spaces. Ethics-wise learners are therefore mindful of the source and state of knowledge, treating AI-generated outputs as provisional inputs, subject to scrutiny, revision, and collective evaluation. Members of the classroom community shoulder collective responsibility for advancing and gatekeeping their collective knowledge (Scardamalia, 2002; Zhang et al., 2009): maintaining the integrity of knowledge-building processes, disagreeing respectfully, recognizing areas where evidence is weak, resisting premature conclusions, and limiting AI use when it threatens student agency or well-being.

Therefore, ethical stewardship of AI is not solely an individual responsibility. Ethical risks are socio-technical, arising from data practices, model behavior, institutional incentives (efficiency, performance pressure), and governance gaps. Educational systems therefore need to work with other partners to develop *system-level stewardship*, including ethics-by-design approaches embedded in educational tools and norms, transparency and traceability mechanisms, and risk-based governance that anticipates, monitors, and mitigates harm (European Commission, 2024).

### *Self- and Community-Growing*

The preceding four principles — being knowledge-wise, intelligence-wise, context-wise, and ethics-wise — guide how learners and educators govern intellectual assets in AI-infused environments. The final principle highlights the purpose of such efforts: *self- and community-growing*. While AI systems can optimize task execution, they cannot choose purposes or sustain commitment to long-term goals. It is a fundamental human responsibility to ensure that AI serves meaningful ends. From an educational standpoint, the core purpose of intellectual stewardship is twofold — self-growing and community-growing — and the two are mutually reinforcing. As learners contribute to collective knowledge and well-being, they develop the dispositions, identities, and commitments needed for personal lifelong growth; conversely, self-growth equips learners to advance their community's knowledge and well-being. Students' work with AI is therefore oriented not toward task performance alone, but toward developing essential human capacities, addressing authentic problems, and contributing to the public good.

With *self-growing* as a guiding aim, students and educators attend to whether AI use contributes to their personal development over time. Students ask not only how AI can help them produce better assignments, but how the process can make them better writers, thinkers, and learners. Teachers do not merely use AI to generate lesson plans; they strive to grow as reflective designers who deepen their understanding of content, attend to students' diverse ideas, and respond in ways that invite deeper thinking. Over time, self-growing stewardship becomes a form of identity development — the ongoing process through which learners negotiate who they are and who they are becoming in relation to knowledge, practice, and community (Carlone & Mercier, 2026). Central to this is cultivating deeper self-knowledge: understanding one's values, experiences, strengths, limitations, and long-term purposes. Learners must cultivate a deeper understanding of themselves as they interact with AI and collaborate with others: What do I value and care about? What am I good at, and how can I further develop these strengths? What are my limitations and struggles, and how should I manage or address them? With a clearer sense of self and long-term purpose, learners can put AI to work to pursue self-sustained intentional learning and continually grow adaptive expertise (Barron, 2006; Bereiter & Scardamalia, 1993; Schwartz, 2025). In AI-infused environments where answers are readily available and effort can be easily bypassed, the dispositions of growth mindset, grit, and perseverance become especially valuable for resisting short-term gratification and invest in deep learning (Duckworth et al., 2007; Dweck et al., 2014). As AI helps reduce the time and effort required for routine cognitive tasks, students may reinvest their resources in addressing greater complexity: working with richer representations, revisiting enduring problems at higher levels, identifying new or deeper issues, and launching new directions of learning (Bereiter & Scardamalia, 1993).

With *community-growing* as a related aim, students consider how their knowledge work with AI advances shared understanding, addresses authentic challenges, and enhances collective good and well-being. In doing so, they govern not only knowledge and ideas but also the social, epistemic, and emotional conditions under which hybrid human–AI knowledge work unfolds. From a Knowledge Building perspective, learning is not merely an individual achievement but a contribution to the advancement of community knowledge—knowledge that is shared, improvable, and oriented toward the public good (Scardamalia & Bereiter, 2014). Community-growing stewardship thus positions students as responsible contributors to the intellectual and social capital of their learning communities and the larger society.

Within their immediate learning communities, students contribute to collective understanding, discourse, inquiry practices, and socio-emotional well-being. They may use AI tools to synthesize collective progress, identify connections among ideas, evaluate community knowledge in relation to peer or global knowledge, highlight promising inquiry directions, or surface social and emotional needs that warrant attention (Bereiter & Scardamalia, 2025; Chen et al., 2023; Feng, 2025). AI-based translation and multimodal representations may also be leveraged to overcome communication barriers and help members with different cultural, linguistic, and knowledge backgrounds engage in collaborative discourse. Beyond their own classrooms, students contribute knowledge to address the needs of broader communities and fields. AI-enhanced designs may support students' sharing of productive knowledge advances across classrooms, enabling ideas to travel, connect, and improve through broader discourse networks (Yuan, Zhang, & Chen, 2022; Zhang et al., 2020). Digital platforms may be designed to support youth engagement in community-based research, such as using public data to investigate locally meaningful issues (Harris et al., 2020; Magnussen & Hod, 2023). In our ongoing design experiment, students collaborate with external organizations to address authentic challenges posed by community partners (Underwood et al., 2025). Students generate guiding questions, collect and analyze data, and refine ideas of value to their partners' problem solving. AI tools can further extend these possibilities by supporting data analysis, synthesis across sources, and communication with diverse audiences.

Across both levels of participation, a key practice for community-growing stewardship is *metadiscourse* — meta-talk through which students reflect on shared goals, collaboration roles, group norms, and progress (Bereiter & Scardamalia, 2016; Zhang et al., 2018). Metadiscourse provides a means for socially shared regulation of cognitive and socioemotional processes in collaborative learning communities (Hadwin et al., 2018; Vokatis et al., 2026). AI-supported analytics may help students navigate metadiscourse: reflecting on the promisingness of ideas, tracing collective knowledge progress, identifying productive patterns, generating rise-above artifacts, and connecting their inquiry with peer classrooms or broader communities (Bereiter & Scardamalia, 2025; Chen et al., 2015; Yuan et al., 2022; Yang et al., 2024; Zhang et al., 2018; Zhu et al., 2022).

## Discussion: Putting the Principles Together

The principles of intellectual stewardship provide a nuanced lens for viewing the changing functions of human minds in AI-infused knowledge environments, foregrounding epistemic agency and reflective judgment to co-construct adaptive learning processes and knowledge practices in an AI-infused world (Pendleton-Jullian & Brown, 2018; Tao & Zhang, 2021; Zhang et al., 2022). As AI takes on various cognitive operations and amplifies potential risks, human minds need to increase their capacity for meta-level judgments and governance. Each principle highlights meta-level stances and capacities that are essential to the functioning of human minds to approach the conceptual, epistemic, social/contextual, and ethical issues of knowledge work with AI. Students enact high-level epistemic agency and responsibility by acting on agentic stances and dispositions, as shown through how they approach the critical decision points aligned with different facets of intellectual stewardship.

- Knowledge-wise stewardship emphasizes the essential role of epistemic understanding and meta-knowledge in AI environments: treating knowledge as improvable, engaging in sustained inquiry, and taking responsibility for advancing the coherence, explanatory power, and public value of ideas—whether generated by humans, AI, or hybrid systems.
- Intelligence-wise stewardship guides how thinking is organized and distributed, highlighting the need for meta-intelligence to govern the use of intellectual processes and tools, so that human judgment, agency, and metacognitive control remain central.
- The context-wise mind embraces sensitivity to place, time, people, and evolving conditions, enabling learners to discern which problems are worth pursuing, how contributions should be situated, and when to adapt or redirect intellectual effort in response to evolving changes and opportunities.
- Ethics-wise stewardship foregrounds responsibility for the consequences of knowledge work, emphasizing the need to nurture ethical minds, caring hearts and social norms in contexts where intellectual power can be potentially misused.

- Finally, self- and community-growing give all the stewardship practices a meaningful purpose and commitment, ensuring that knowledge work serves personal development, collective well-being, and the public good rather than short-term efficiency and task completion alone.

As a whole, the meta-level dispositions and capabilities articulated by the principles depict a portrait of wisdom-oriented social minds that work adaptively to navigate collaborative knowledge practices with AI for meaningful purposes. These mindsets and practices are characteristic of wise, responsible, and fruitful stewards of intellectual assets and practices. As AI assists various cognitive processes, human minds need to adapt to develop such meta-level dispositions and capabilities that help make wise judgments and responsible choices about core issues of thinking, learning, and knowledge work. In a collaborative community, such dispositions are cultivated as both personal habits of mind and socially shared cultural norms, guiding members' reflective judgments, participatory roles, interactional processes in their knowledge-building work distributed across human and AI systems.

A core conjecture of this framework is that different stances and choices toward AI use do not merely affect student performance in particular learning tasks; they lead to different participatory pathways during the activities and different developmental trajectories in the long term. In light of the principles and related studies, it is likely that students who adopt productive stances toward knowledge and intellectual power will continually create new opportunities for deep thinking, expansive learning and collaborative knowledge building (Chen et al., 2023; Hou et al., 2026; Feng, 2025). As knowledge-wise participants, they treat AI outputs as provisional contributions—objects for examination, revision, and further inquiry—rather than finished answers. As intelligence-wise actors, they choose thought-demanding engagement over cognitive shortcuts, deliberately regulating when AI comes in to extend their thinking and when deeper personal effort is required. As context-wise learners, they situate their work in relation to authentic problems and evolving knowledge needs, reading the broader intellectual and social landscape rather than focusing narrowly on assignment completion. As ethics-wise learners, they take responsibility for advancing collective good through their knowledge work, attending to issues of accuracy, integrity, fairness, and safety. And with self- and community-growing as their guiding purposes, they reinvest efficiency gains into deeper exploration, greater complexity, and more meaningful contributions. Over time, this will position students on self-sustained, ever-expanding learning trajectories, supporting the development of adaptive expertise, epistemic agency, and personal identities grounded in responsibility and contribution to the public good (Carlone & Mercier, 2026; Engle et al., 2012; Scardamalia & Bereiter, 2014; Schwartz, 2025).

By contrast, unproductive stewardship reduces AI use to efficiency and performance optimization. Knowledge work becomes synonymous with task completion rather than sustained idea advancement. AI is used to generate answers, polish outputs, or bypass difficulty, fostering cognitive offloading without deep understanding (Kasneci et al., 2023; Kosmyna et al., 2025; Shen & Tamkin, 2026). Over time, student intellectual engagement may decline and their epistemic agency may weaken. Learners may prioritize effort-saving over sense-making, external performance over intellectual inquiry, and superficial compliance over intentional participation and contribution. In more troubling cases, AI becomes a mechanism for academic disengagement or dishonesty, undermining both personal development and collective well-being. Future research must systematically explore these possibilities and create principled designs for AI-infused learning environments that expand productive learning trajectories for all students.

## Practical Implications and Future Research

The framework of intellectual stewardship provides theory-based guidance for advancing creative knowledge practices amid the opportunities and risks of AI. As AI augments or automates cognitive tasks, schools must refocus education on cultivating wise and responsible intellectual stewards — learners, workers, and citizens who are knowledge-wise, intelligence-wise, context-wise, and ethics-wise, and who purposefully seek continual self- and community-growing. Beyond learning to use AI tools efficiently, learners need authentic knowledge-building environments that offer rich opportunities to practice wise judgment about their intellectual work: deciding when, how, and why AI should be used — and when it should not. The principles of intellectual stewardship highlight the reflective questions that guide such choice-making and inform teacher design and reflection. Table 1 offers example questions aligned with each principle as practical scaffolds for learner self-regulation and teacher planning.

Table 1: Reflective questions for learners and teachers as intellectual stewards

| Principle | Reflection points for learners | Reflection points for teachers |
|---|---|---|
| **Knowledge-Wise:** Understanding the status of knowledge and taking fruitful actions | What am I trying to understand or achieve? Which problems are worth pursuing, and which ideas or information matter? What is the current state of our knowledge? What problems, gaps, or tensions remain? How does this idea improve on what we already know? How can we further improve or rise above our ideas (with AI or not), toward what aims? | What kinds of ideas and problems are worth sustained inquiry here? How does this activity support collective idea improvement rather than task completion? How will students know whether knowledge has actually advanced? How can I support their ongoing reflection and meta-discourse? |
| **Intelligence-Wise:** Orchestrating intellectual processes and resources | What kinds of thinking or inquiry are needed here? How should it be carried out, by/with whom, using what intellectual resources and tools? Which human and AI contributions are appropriate? What counts as good or success? How should I balance AI automation with my own reasoning, thinking and judgment? | What forms of thinking should remain primarily human, and where can AI productively contribute? How can AI be used to extend—not replace—students' reasoning, creativity, and judgment? How will students learn to think with AI output and avoid overreliance on AI? |
| **Context-Wise:** Situated discernment and responsiveness in a changing environment | What is happening in this context that calls for intellectual engagement? Why does this work matter here and now? How should I position my learning and contribution in relation to the works and ideas of others? | What real problems, contexts, or communities can anchor this work meaningfully? How can I design environments that help students read emerging needs, opportunities, and tensions? How should the design adapt as inquiry directions evolve? |
| **Ethics-Wise:** Exercising moral judgment, | What responsibilities accompany my learning and work? What is right or good to do, and what is not? Should this work be done, and under what conditions? | What ethical risks might arise in this learning activity (e.g., authorship, bias, responsibility, integrity, misuse)? What norms, |

| accountability and care | Who might be affected by these ideas or outputs? How do I ensure integrity, fairness, and accountability? | routines, or checkpoints can help students practice ethical judgment? When should AI use be constrained or slowed down? |
|---|---|---|
| **Self- & Community-Growing:** Human-centered purpose, commitment, and contribution | What kind of person do I want to be? How does this work help me grow as a learner, thinker, writer and person? What responsibilities do I have to my learning community and the broader society? How does my work advance our shared understanding or collective well-being? | How does this learning experience support students' long-term growth, identity development, and sense of agency? How does it strengthen the intellectual and social dynamics of the learning community? What opportunities exist for students to contribute beyond the classroom? |

Future research should further develop understanding of the reflective stances and judgments associated with the five principles; trace how these personal and shared stances interact to shape learners' engagement with human and AI systems over time; and examine their consequences for collaborative knowledge building, epistemic agency, identity development, and adaptive expertise. Design-based research could investigate ways to scaffold human-AI interaction in knowledge-building environments and assess impacts on students' conceptual understanding, collaborative discourse, and evolving sense of responsibility for knowledge advancement. Insights from such work will inform the human-centered design of AI systems that scaffold principle-based judgment and collective responsibility. Beyond AI tutors that deliver personalized instruction of predefined content, education needs what Resnick (2024) calls "a truly personal approach to learning" — one that gives learners choice and control: to set their own goals, develop their own ideas and strategies, and reflect on their progress over time. AI-infused environments should support such self-sustained learning, keeping learners deeply engaged in authentic challenges, monitoring progress, and purposefully seeking higher-level goals and challenges.

Future work should also extend the intellectual stewardship framework to teachers, school leaders, policymakers, and community partners. Creating a new learning culture in the age of AI requires all stakeholders to ensure that human wisdom, creativity, and responsibility for social good continually rise alongside the expansion of technological power. In this collaborative adventure, we are all called to adapt and re-adapt as wise, responsible, and fruitful stewards.

## Acknowledgements

The writing of this paper was supported in part by the Faculty Innovation Fellowship program of the University at Albany, State University of New York. In the writing process, I used Microsoft 365 Copilot, ChatGPT, and Claude as AI writing assistants to search for related literature, explore phrasing of key concepts, and proofread drafts for clarity and formatting. The substance and ideas of the paper are entirely my own.